\documentclass[
    ,final            
  ]
  {aipproc}

\layoutstyle{6x9}

\begin{document}

\def\D{\displaystyle}
\def\r{\rho}
\def\R#1{\rho^{(#1)}}
\def\H#1{{\cal H}^{\otimes #1}}
\def\tr{{\rm tr}}
\def\boldp{{\bf p}}
\def\boldf{{\bf f}}
\def\boldn{{\bf n}}
\def\sx{\sigma_x}
\def\sy{\sigma_y}
\def\sz{\sigma_z}
\def\rxyz{\rho_{x,y,z}}
\def\intdr{\int\!\! d\rho\,}

\def\ket#1{|#1\rangle}
\def\bra#1{\langle #1|}

\title{Priors in quantum Bayesian inference}


\author{Christopher A. Fuchs}{
  address={Perimeter Institute for Theoretical Physics,
31 Caroline Street North,
Waterloo, Ontario, Canada N2L 2Y5}
}

\author{R\"udiger Schack}{
  address={Department of Mathematics,
           Royal Holloway, University of London, Egham, Surrey TW20 0EX, UK}
}

\begin{abstract}
  In quantum Bayesian inference problems, any conclusions drawn from a finite
  number of measurements depend not only on the outcomes of the measurements
  but also on a prior. Here we show that, in general, the prior
  remains important even in the limit of an infinite number of measurements.
  We illustrate this point with several examples where two priors lead
  to very different conclusions given the same measurement data.
\end{abstract}

\maketitle

\section{Introduction}

This paper addresses the problem of inference in quantum mechanics in a very
general setting. Consider a sequence of quantum systems, each with Hilbert
space ${\cal H}$ which, for simplicity, is assumed to have finite dimension.
Now some  of the systems are measured. What conclusions can be
drawn from the measurement outcomes?

It is often useful to consider less general situations. One of the most
studied quantum inference problems is quantum state estimation, which is
frequently described as follows \cite{Rehacek2004}. Each system is assumed to
have the same unknown state $\sigma$, where $\sigma$ is a density operator on
${\cal H}$. A sequence of appropriately chosen measurements is then used to
determine $\sigma$. The most general single-system measurement is described by
a POVM, which is a collection of positive operators, $\{E_1,\ldots,E_r\}$,
acting on ${\cal H}$, such that $\sum_k E_k$ is the identity operator. The
index $k\in\{1,\ldots,r\}$ labels the possible measurement outcomes. If the
state of the system is $\rho$, the probability of obtaining outcome $k$ is
given by $\tr(\rho E_k)$.

The Bayesian solution to the quantum state estimation problem is
straightforward. Its starting point is a prior probability distribution
$p_{\rm prior}(\rho)$ on the set of density operators on ${\cal H}$.  Assume
that the POVM $\{E_k\}$ is measured for a single system. If the outcome of the
measurement is $k$, the distribution $p_{\rm prior}(\rho)$ is updated using a
quantum version of Bayes's rule \cite{Brun2001a,Bernardo1994},
resulting in the posterior distribution
\begin{equation} \label{qbayes}
p_{\rm posterior}(\rho)=\frac{p_{\rm prior}(\rho)\tr(\rho E_k)}{\int d\rho\,p_{\rm
    prior}(\rho)\tr(\rho E_k)} \;.
\end{equation}
This process is iterated, each time using a fresh copy of the system and
possibly a different measurement, and each time setting $p_{\rm prior}(\rho)$
equal to the previously obtained $p_{\rm posterior}(\rho)$. Given some mild
conditions on the original prior distribution, and assuming that the
measurements are appropriately chosen, it can be shown that the iterated
distribution approaches a delta distribution centered at the unknown state
$\sigma$. For instance, this convergence holds if the original $p_{\rm
  prior}(\rho)$ is nonzero for all $\rho$ and the measured POVM is
informationally complete \cite{Caves2002b}.

In practice, the iteration will have to stop after a finite number of
measurements. The posterior distribution will then generally depend on the
initial prior distribution. From a Bayesian perspective, this dependence on
the prior distribution is unavoidable. Finding the appropriate mathematical
form of the prior distribution is a central part of the Bayesian approach to
quantum state estimation.

There are a number of (non-Bayesian) state estimation methods that attempt to
circumvent the dependence of the estimate on the prior distribution
\cite{Rehacek2004}. An important example is maximum likelihood estimation. We
are not going to discuss non-Bayesian methods any further in this paper.

In the next section, we will show that quantum state estimation as described
above is a special case of a far more general quantum inference scenario in
which priors retain their importance even in the limit of an
infinite number of measurements. This is followed by a section illustrating
this point with some examples. The paper concludes with a brief discussion.

\section{General quantum inference problem}

There is something peculiar about the quantum Bayes rule (\ref{qbayes}).
Since a density operator $\rho$ encodes the probabilities for outcomes of
quantum measurements, the rule (\ref{qbayes}) mixes two kinds of
probabilities, namely the ``classical'' probabilities $p_{\rm prior}$ and
$p_{\rm posterior}$ on the one hand, and the ``quantum'' probabilities encoded
in $\rho$ on the other hand. We will now show that the rule (\ref{qbayes}) is
a special case of a more general rule phrased entirely in terms of quantum
states. It is therefore not necessary to make the distinction between two
kinds of probability.

There is a general updating rule built into quantum theory. To describe a
quantum measurement fully, one needs to provide, in addition to the POVM
$\{E_1,\ldots,E_r\}$, a set of {\em Kraus operators\/} $A_{kj}$ such that
$E_k=\sum_j A_{kj}A_{kj}^\dagger$ for $k=1,\ldots,r$ \cite{Kraus1983}. If the
measurement gives outcome $k$ for a system in state $\rho$, the state of the
system after the measurement will be
\begin{equation} \label{kraus}
\rho_k = \tr(\rho E_k)^{-1}\sum_j A_{kj} \rho A_{kj}^\dagger \;.
\end{equation}
This rule has an interpretation very similar to the Bayes rule. If we call
$\rho$ the prior state and $\rho_k$ the posterior state, we see that the prior
state is changed into the posterior state upon acquisition of the data $k$.

To see that the quantum Bayes rule (\ref{qbayes}) is a special case of the
Kraus rule (\ref{kraus}) \cite{Brun2001a}, consider a sequence of quantum
systems each with Hilbert space ${\cal H}$ as before. We define a {\em
  prior\/} on our sequence of systems as a sequence of $n$-system states
$\rho_{\rm prior}^{(n)}$, $n=1,2,\ldots$, where each $\rho_{\rm prior}^{(n)}$
is a density operator on the $n$-fold tensor-product Hilbert space ${\cal
  H}^{\otimes n}={\cal H}\otimes\cdots\otimes{\cal H}$, and where $\rho_{\rm
  prior}^{(n)}=\tr_{n+1}\rho_{\rm prior}^{(n+1)}$ for all $n\ge1$. By
$\tr_{n+1}$ we denote the trace over the $(n+1)$-th system. In words, each
member of the sequence is obtained from the next by tracing over the
additional system.

Now assume that the first system is measured and an outcome $k$ is
obtained. Applying the rule (\ref{kraus}) we find, for any
$n\ge1$, that the state of the first $n$ systems after the measurement is
\begin{equation}
\rho_k^{(n)} =
\tr(\rho_{\rm prior}^{(n)} E_k)^{-1}
\sum_j A_{kj} \rho_{\rm prior}^{(n)} A_{kj}^\dagger \;,
\end{equation}
where it is understood that the operators $A_{kj}$ and $E_k$
act on the first system
only. By tracing over the first system, we obtain what we call the
{\em posterior\/} on our sequence of systems,
\begin{equation}
\rho_{\rm posterior}^{(n)} = \tr_1 \rho_k^{(n+1)} \;\;\;(n=1,2,\ldots)\;.
\end{equation}
The posterior is again a sequence of states; its $n$-th member is obtained
from the $(n+1)$-th member of the prior by measuring and then discarding the
first system. The posterior has the property $\rho_{\rm posterior}^{(n)} =
\tr_{n+1}\rho_{\rm posterior}^{(n+1)}$ for all $n\ge1$ and thus has the form
of a prior. We can therefore iterate the above procedure, each time setting
the prior equal to the posterior obtained in the previous iteration.

Given the prior and the measurement data, the posterior is the unique sequence
of states for the remaining (i.e., not yet measured) systems. In this sense the
posterior constitutes the unique correct solution of the quantum inference
problem. In particular, the one-system state $\rho_{\rm posterior}^{(1)}$ is
the marginal state for the first unmeasured system. This state is sometimes
called the Bayesian mean estimator.

To recover the familiar rule (\ref{qbayes}), only one simple additional
assumption has to made, namely that for any $n\ge1$ the state $\rho_{\rm
  prior}^{(n)}$ is symmetric under permutations of the $n$ systems. In this
case we say that the prior is {\em exchangeable} \cite{Hudson1976}. Given this
extra assumption, it is the content of the quantum de Finetti theorem
\cite{Caves2002b,Hudson1976} that the prior can be written as
\begin{equation} \label{exchangeablePrior}
\rho_{\rm prior}^{(n)} = \int d\rho\,p_{\rm prior}(\rho)\,\rho^{\otimes n}
\;\;\;(n=1,2,\ldots)\;,
\end{equation}
where $p_{\rm prior}(\rho)$ is a probability distribution on the space of
single-system density operators, and $\rho^{\otimes n}$ is the $n$-fold tensor
product $\rho\otimes\cdots\otimes\rho$. It is not difficult to establish
\cite{Brun2001a} that a measurement on the first system with outcome $k$ will
lead to the posterior
\begin{equation}
\rho_{\rm posterior}^{(n)} = \int d\rho\,p_{\rm  posterior}(\rho)\,\rho^{\otimes n}
\;\;\;(n=1,2,\ldots)\;,
\end{equation}
with $p_{\rm posterior}(\rho)$ given by the rule (\ref{qbayes}).

As was pointed out in the introduction, in the limit of an infinite number of
iterations, $p_{\rm posterior}(\rho)$ typically approaches a delta function
which is independent of the detailed functional form of $p_{\rm
  prior}(\rho)$. This conclusion, however, depends crucially on the
assumption that the prior is of the form (\ref{exchangeablePrior}), i.e., that
the prior is exchangeable. In other words, even in the limit of an infinite
number of iterations, conclusions depend on the prior. All we can say is that
some details of the prior become irrelevant in this limit.

Exchangeable priors are an important class of priors that are used so
frequently that it is sometimes overlooked that exchangeability is an
assumption.  Making this assumption is equivalent to choosing a prior from a
restricted set. What we have therefore shown in this section is that
conclusions drawn in Bayesian quantum inference situations generally depend on
the prior as well as measurement data, even in the limit of an infinite number
of measurements.  In the next section, we illustrate this point with some
examples.

\section{Example priors}

The first example is a sequence of qubits, i.e., two-dimensional quantum
systems. We denote by $\ket0$ and $\ket1$ two orthogonal basis states and
consider three priors.

Our first prior is exchangeable and given by
\begin{equation} \label{purePrior}
\rho_a^{(n)} = \int d\ket\psi\,p_a(\ket\psi)\;(\ket\psi\bra\psi)^{\otimes n}
\;\;\;\;(n=1,2,\ldots)\;,
\end{equation}
where $d\ket\psi\,p_a(\ket\psi)$ is the Haar measure on the space of pure
one-qubit states.

Our second prior is also exchangeable, but consists of a sequence of pure
product states. It could be called a {\em Rosenkrantz and Guildenstern
  prior\/} \cite{Stoppard1967}. This is the state one might assign to a
quantum random number generator manufactured by a trusted company. It is given
by
\begin{equation} \label{RosenkrantzPrior}
\rho_b^{(n)} = \Big(\frac12(\ket0+\ket1)(\bra0+\bra1)\Big)^{\otimes n}
\;\;\;(n=1,2,\ldots)\;.
\end{equation}

Our third prior is not exchangeable. We call it a {\em
  counter-inductive} prior \cite{Smolin2006} because it leads
one to predict outcomes that are the opposite of what an argument by induction
would suggest. In particular, we will see that updating this prior after a
string of $m$ identical measurement outcomes, the probability
for obtaining the opposite outcome in the next measurement approaches 1 as
$m$ increases. The counter-inductive prior is given by
\begin{eqnarray} \label{counterinductivePrior}
\rho_c^{(n)}  = & & {\cal N} \Big[\; \sum_{k=1}^{n-1} 2^{-k^2}\Big(\ket0\bra0^{\otimes
    k}\otimes\ket1\bra1^{\otimes(n-k)}
 + \ket1\bra1^{\otimes k}\otimes\ket0\bra0^{\otimes(n-k)}\Big) \nonumber\\
 & & +  \sum_{k=n}^\infty 2^{-k^2}
       \Big(\ket0\bra0^{\otimes n}+\ket1\bra1^{\otimes n}\Big)\;\Big]
\;\;\;\;\;\;(n=1,2,\ldots)\;,
\end{eqnarray}
where the normalization constant ${\cal N}$ is determined by the equation
\begin{equation}
1 = 2 {\cal N} \sum_{k=1}^\infty 2^{-k^2} \;.
\end{equation}
It is not difficult to check that this sequence of states satisfies the
defining condition of a prior, $\rho_c^{(n)}=\tr_{n+1}\rho_c^{(n+1)}$ for $n\ge1$.

We now imagine that a sequence of von Neumann measurements in the
$\{\ket0,\ket1\}$ basis is carried out and that each measurement produces the
same outcome, 0. The measurement data thus consist of a string of
zeros. These data are used to update iteratively each of our three priors
above. For each prior, we  compute the marginal one-system posteriors in
the limit of an infinite number of iterations.

For the two exchangeable priors, we obtain easily the limits
\begin{equation}
\rho_a^{(1)} \to \ket0\bra0 \;\;\mbox{ (number of iterations $\to\infty$)}
\end{equation}
and
\begin{equation}
\rho_b^{(1)} \to \frac12(\ket0+\ket1)(\bra0+\bra1)
\;\;\mbox{ (number of iterations $\to\infty$)}.
\end{equation}
To compute the limit for the counter-inductive prior $\rho_c^{(n)}$, we first
compute the posterior after $m$ iterations, $\rho_{c,m}^{(n)}$. We find, for
$n=1,2,\ldots$,
\begin{equation}
\rho_{c,m}^{(n)}  = {\cal N}_m \Big(\; \ket1\bra1^{\otimes n}
+ \sum_{k=1}^{n-1}  2^{-k(2m+k)}\ket0\bra0^{\otimes k}\otimes\ket1\bra1^{\otimes(n-k)}
+ \sum_{k=n}^\infty 2^{-k(2m+k)}\ket0\bra0^{\otimes n} \;\Big)\;,
\end{equation}
where ${\cal N}_m$ is determined by the condition
\begin{equation}
1 = {\cal N}_m \sum_{k=0}^\infty 2^{-k(2m+k)} \;,
\end{equation}
implying that $\D{\lim_{m\to\infty}{\cal N}_m=1}$. Hence
\begin{equation}
\rho_c^{(1)} \to \ket1\bra1 \;\;\mbox{ (number of iterations $\to\infty$)}.
\end{equation}

Clearly, the three priors lead to radically different conclusions for the same
infinite sequence of data.

We now move on to our second example, where we compare two priors, again for a
sequence of qubits. For our first prior, we choose a generic exchangeable
prior,
\begin{equation}
\rho_d^{(n)} = \int d\rho\,p_d(\rho)\,\rho^{\otimes n}
\;\;\;(n=1,2,\ldots)\;,
\end{equation}
where $d\rho\,p_d(\rho)$ is a measure on one-qubit density operators,
i.e., density operators on a two-dimensional Hilbert space, ${\cal H}$. We
assume that $p_d(\rho)$ is nonzero for all $\rho$ on ${\cal H}$. This prior
entails that there is no entanglement between the qubits.

By contrast, our second prior, though identical with our first on the single system marginals, does not rule out entanglement between pairs of
qubits. For even numbers of systems, it is defined by
\begin{equation} \label{ePriorEven}
\rho_e^{(2n)} = \int d\sigma\,p_e(\sigma)\;\sigma^{\otimes n}
\;\;\;\;(n=1,2,\ldots)\;,
\end{equation}
where $d\sigma\,p_e(\sigma)$ is a measure on the space of two-qubit density
operators, i.e., density operators on the four-dimensional Hilbert space
${\cal H}\otimes{\cal H}$.  For odd numbers of systems, the prior is defined
by
\begin{equation} \label{ePriorOdd}
\rho_e^{(2n-1)} = \tr_{2n}\rho_e^{(2n)} \;\;\;\;(n=1,2,\ldots)\;.
\end{equation}

Assume now that a sequence of informationally complete measurements is
performed on the sequence of qubits. As before, the measurement data are used
to update both our priors iteratively. We assume that the data are such that
for the second prior, the marginal two-system posterior converges to an
entangled two-qubit state, e.g., the maximally entangled state
$\rho_{ME}={\frac12}(\ket{00}+\ket{11})(\bra{00}+\bra{11})$. We thus assume
that
\begin{equation}
\rho_e^{(2)} \to \rho_{ME} \;\;\mbox{ (number of iterations $\to\infty$)}.
\end{equation}
Since for the maximally entangled state, both marginal states are equal to the
totally mixed state $\rho_{M}=\frac12(\ket0\bra0+\ket1\bra1)$, it follows that
given the same data, the marginal one-system posterior for our first prior
converges to $\rho_{M}$,
\begin{equation}
\rho_d^{(1)} \to \rho_{M} \;\;\mbox{ (number of iterations $\to\infty$)},
\end{equation}
and the marginal two-system posterior converges to
\begin{equation}
\rho_d^{(2)} \to \rho_{M}\otimes\rho_M \;\;\mbox{ (number of iterations $\to\infty$)},
\end{equation}
which is equal to the maximally mixed state of two qubits and, of course, not
entangled.

Once more, we see that the same infinite sequence of data leads to
radically different conclusions for the two priors.

\section{Conclusion}

The most general way in quantum mechanics for obtaining a quantum state from
data is via the Kraus rule (\ref{kraus}). It is clear that the quantum state
obtained in this way generally depends on some prior state in addition to the
data. In an earlier paper \cite{Caves2007} we have established the general
principle that a quantum state is never determined by measurement data alone.
This is true even in state preparation, because the prepared state always
depends on the prior quantum state of the preparation device \cite{Caves2007}.

What we have illustrated here it that this general principle continues to hold
in situations where measurements are repeated many times. A quantum state is
never determined by measurement data alone, even in the limit of infinitely
many measurements.

\end{document}